\documentstyle[twocolumn,aps]{revtex}
\input{epsf}
\begin{document}
\title{Hamiltonian self-adjoint extensions for (2+1)-dimensional Dirac
particles}
\author{H. Falomir and P.\ A.\ G.\ Pisani}
\address{IFLP - Departamento de F\'\i sica, Fac.\ de Ciencias Exactas, UNLP,
C.C.\ 67, (1900) La Plata, Argentina}
\date{\today}
\maketitle
\begin{abstract}

Abstract: We study the stationary problem of a charged Dirac
particle in (2+1)-dimensions in the presence of a uniform
magnetic field $B$ and a singular magnetic tube of flux $\Phi = 2
\pi \kappa/e$. The rotational invariance of this configuration
implies that the subspaces of definite angular momentum $l+1/2$
are invariant under the action of the Hamiltonian $H$. We show
that, for $\kappa-l\geq 1$ or $\kappa-l\leq 0$, the restriction of
$H$ to these subspaces, $H_l$, is essentially self-adjoint, while
for $0<\kappa-l<1$, $H_l$ admits a one-parameter family of
self-adjoint extensions (SAE). In the later case, the functions
in the domain of $H_l$ are singular (but square-integrable) at
the origin, their behavior being dictated by the value of the
parameter $\gamma$ that identifies the SAE. We also determine the
spectrum of the Hamiltonian as a function of $\kappa$ and
$\gamma$, as well as its closure.

\end{abstract}
\pacs{}

\narrowtext

\section{Introduction}

In Quantum Mechanics, observables are realized in terms of
self-adjoint operators on a Hilbert space. It is for these
operators that the spectral theorem holds \cite{R-S}. In
particular, the dynamics of a quantum system should be given by a
unitary group whose generator, the Hamiltonian $H$ (usually a
differential operator acting on an appropriate space of square
integrable functions), must be self-adjoint.

In general, physical considerations lead to a {\it formal}
expression for the Hamiltonian, although they can leave its
domain of definition not completely specified. Usually, one can
choose a dense subspace of the Hilbert space on which $H$ is
well-defined and {\it symmetric}, but not necessarily {\it
self-adjoint}.

In these conditions, the question is posed of determining if the
expression found for $H$ has a unique self-adjoint extension in
the Hilbert space (i.e., if $H$ is {\it essentially
self-adjoint}), or it admits different self-adjoint extensions
(SAE) (differing in the physics they describe) and, in this case,
which one corresponds to the physical system under consideration.

\bigskip

A situation of practical interest in which the Hamiltonian admits
nontrivial self-adjoint extensions corresponds to the movement of
charged particles under the influence of a Bohm - Aharonov
singular magnetic flux tube \cite{A-B}, like fermions in the
presence of cosmic strings \cite{deSG} or non-relativistic
spinless quantum particles interacting with a thin solenoid
\cite{A-T}. In references \cite{deSG,A-T,S,S1}, this problem has
been analyzed by means of von Neumann's theory of deficiency
subspaces \cite{R-S}.

This kind of situations have also been studied as a limit of a
smeared flux, using a $\delta$-function shell magnetic field
\cite{H,A et al} or uniform magnetic fields confined to a finite
tube \cite{T,C}, and a punctured plane \cite{Poly,BDS}, which
leads to the consideration of boundary conditions at a finite
radius, both spectral and local. It has also been of interest the
study of charged particle states bounded to flux tubes
\cite{Bento,Bordag,Cava1,Cava2,Moroz}.

The presence of a $\delta$-like magnetic field has also been
considered in connection with vacuum polarization effects in
\cite{F-L}, to model the presence of a point-like impurity in a
bidimensional system \cite{D-O-T}, and more recently to describe
a non-relativistic electron in the presence of a uniform
electromagnetic field and a singular vortex, as a step toward its
application to the quantum Hall effect \cite{G-S}. This
configuration can also be relevant to the description of
quasiparticles in unconventional superconductors
\cite{Tesa1,Tesa2}.

\bigskip

It is the aim of this paper to study the behavior of a Dirac
electron of mass $M$ and charge $e$ constrained to live in a
(2+1)-dimensional space, in the presence of a constant magnetic
field $B$ and a singular magnetic flux tube $\Phi = 2 \pi
\kappa/e$ passing through the origin. In so doing, we will use
von Neumann's theory of deficiency indices to determine the
existence of nontrivial self-adjoint extensions for the
Hamiltonian, a problem that, as far as we know, has not yet been
solved.

The rotational symmetry of the problem allows for studying the
action of the Hamiltonian (a differential operator $H$ defined on
an appropriately restricted set of smooth functions) in each
invariant subspace characterized by a definite angular momentum
$l+1/2$. We find that the restriction of $H$ to the subspaces with
$\kappa-l\geq 1$ or $\kappa-l\leq 0$, $H_l$, is essentially
self-adjoint, while for $0<\kappa-l<1$ the operator $H_l$ admits
a one-parameter family of self-adjoint extensions. In the later
case, the functions in the extended domain of $H_l$ become
singular (though square-integrable) at the origin, their behavior
being dictated by the value of the parameter $\gamma$ that
identifies the SAE.

Finally, we also determine the spectrum of the Hamiltonian as a
function of $\kappa$ and $\gamma$.

\section{Formulation of the problem}

Let us consider a Dirac particle of mass $M$ and charge $e$ in a
$2+1$-dimensional spacetime, in the presence of a uniform
magnetic field $B$ and a singular magnetic flux tube $\Phi = 2 \pi
\kappa/e$ passing through the origin (i.e.\ the flux originated in
a magnetic field which is null at each point of the plane except
at the origin, and whose flux through every curve enclosing the
origin is finite.)

The wave function of this particle is a two component spinor
$\psi$ satisfying the Dirac equation (we adopt the fundamental
units for which $\hbar=1=c$),
\begin{equation}
(i\not \! \! D-M)\psi=0,
\end{equation}
where the covariant derivative is $\not \! \! D=\not
\!\partial-ie\not \! \!A$\footnote{\label{gammas} We choose the
following representation of the $\gamma$-matrices:
\begin{equation}\label{gam-mat}
\gamma^{0}=\sigma^{3}, \ \gamma^{1}= -i\sigma^{2}, \
\gamma^{2}=i\sigma^{1},
\end{equation}
where the $\sigma^{i}, \ i=1,2,3$, are the Pauli matrices. In a
3-dimensional space-time, a non-equivalent representation is
obtained by changing the sign of the matrices, $\gamma^\mu
\rightarrow - \gamma^\mu$, but this amounts to changing the sign
of the parameter $M$, which therefore can be considered to take
real values.}.

We choose the following expression for the vector potential
leading to the magnetic field under consideration,
\begin{equation}
  \vec{A}=\left(\frac{\Omega r}{e} +
  \frac{\kappa}{e r} \right)\hat{e}_\theta,
\end{equation}
where $\Omega=eB/2$ has units of squared mass and $\hat{e}_\theta$
is the unit vector orthogonal to the radial direction.

Accordingly, we get for the Dirac Hamiltonian
$H_{D}=\sqrt{\Omega}H$, where $H$ is the dimensionless
differential operator
\[ H= \]
\begin{equation}\label{hd}
 \left[\begin{array}{cc}
    m & ie^{-i\theta}
    (\partial_{x}-\frac{i}{x}\partial_{\theta}-x-\frac{\kappa}{x}) \\
    -ie^{i\theta}
    (-\partial_{x}-\frac{i}{x}\partial_{\theta}-x-\frac{\kappa}{x}) & -m \
  \end{array}\right],\nonumber \\
\end{equation}
expressed in polar coordinates $(x=\sqrt{\Omega}r,\theta)$, with
$m=M/\sqrt{\Omega}$, the particle mass in units of $\Omega^{1/2}$.

Since $H$ commutes with the angular momentum operator,
$J=-i\partial_{\theta}+\sigma^{3}/2$, the subspaces spanned by the
two-component spinors of the form
\begin{equation}
  \psi(x,\theta)=\left(\begin{array}{c}
    e^{i l \theta}\phi_{(x)} \\
    e^{i(l+1)\theta}\chi_{(x)} \
  \end{array}\right)\in L_2(\textbf{{R}}^2,x\, dx\,d\theta ), \quad l \in \textbf{Z}
\end{equation}
are left invariant by the action of $H$. The restriction of $H$
to each subspace characterized by $l$, $H_l$, can be cast into the
form
\begin{equation}\label{ham}
  H_{l}=
\left(  \begin{array}{cc}
    m & i\left(\frac{d}{dx}+\frac{1-\alpha}{x}-x\right) \\
    i\left(\frac{d}{dx}+\frac{\alpha}{x}+x\right) & -m \
  \end{array}\right),
\end{equation}
with $\alpha = \kappa -l$, when acting on two-component functions
of the radial coordinate,
\begin{equation}
  \psi(x) = \left(\begin{array}{c}
    \phi_{(x)} \\
    \chi_{(x)} \
  \end{array}\right),
\end{equation}
where $\phi_{(x)},\chi_{(x)}\in L_2(\textbf{{R}}^+,2\pi x\, dx)$.

In order to ensure that $H_l$ be symmetric and well-defined we
can restrict its domain to
\begin{equation}\label{domHl}
  {\cal D}(H_l)\equiv
C^{\infty}_{0}(\textbf{R}^+),
\end{equation}
the subspace of functions with compact support away from the
origin and continuous derivatives of all order, which is dense in
$L_2(\textbf{{R}}^+,2\pi x\, dx)$.

To determine whether $H_l$ so defined is (essentially)
self-adjoint we must compute its deficiency indices in the
Hilbert space $L_2(R^{+},2\pi\, x \, dx)$, i.e., the dimensions
of the characteristic subspaces ${\cal K}_\pm$ of its adjoint,
$H_l^\dagger$, corresponding to eigenvalues $\pm i$,
\begin{equation}\label{indices}
  n_\pm = {\rm dim}{\cal K}_\pm.
\end{equation}

In the following we shall show that $H_l$ admits self-adjoint
extensions for $0<\kappa-l<1$, being essentially self-adjoint for
the other angular momentum subspaces.

\section{Self-adjoint extensions}\label{Section-SAE}

In order to determine the deficiency indices of the operator $H_l$
defined in the previous section, we must determine the deficiency
subspaces ${\cal K}_\pm $.

Let us recall that the domain of $H_l^\dagger$, ${\cal
D}(H_l^\dagger)$, is the set of functions $f(x) =
\left(\begin{array}{c}
    f_1 (x) \\
    f_2 (x) \
  \end{array}\right) \in L_2(\textbf{{R}}^+,2\pi x\, dx)$,
for which functions $g(x) = \left(\begin{array}{c}
    g_1 (x) \\
    g_2 (x) \
  \end{array}\right) \in L_2(\textbf{{R}}^+,2\pi x\, dx)$ exist, such that
\begin{equation}\label{fun-lin-cont}
  (f,H_l \psi)=(g,\psi)
\end{equation}
for any $\psi \in {\cal D}(H_l)$. The adjoint $H_l^\dagger$ is
defined by $g=H_l^\dagger f$.

Taking into account eq. (\ref{domHl}) and the expression for
$H_l$, eq. (\ref{ham}), one can easily see that, away from the
origin, the first weak derivative of $f(x)$ is locally in
$L_2(\textbf{{R}}^+,2\pi x\ dx)$. Therefore, by Sobolev's lemma
(see Ref.\ \cite{R-S}), $f(x)$ is absolutely continuous. This
allows for an integration by parts in eq. (\ref{fun-lin-cont}),
which gives
\begin{equation}\label{Hldagger}
  \left(\begin{array}{c}
      g_1 \\
      g_2 \
    \end{array}\right)=
    \left(  \begin{array}{cc}
    m & i\left(\frac{d}{dx}+\frac{1-\alpha}{x}-x \right)\\
    i\left( \frac{d}{dx}+\frac{\alpha}{x}+x \right)& -m \
    \end{array}\right)
    \left(\begin{array}{c}
      f_1 \\
      f_2 \
    \end{array}\right).
\end{equation}

In conclusion, $H_l^\dagger$ acts as a differential operator in
the same way as $H_l$ in eq.\ (\ref{ham}), but on a larger domain
${\cal D}(H_l^\dagger) (\supset {\cal D}(H_l))$, consisting of the
subspace of functions of $L_2(\textbf{{R}}^+,2\pi x\, dx)$ which
are absolutely continuous in $\textbf{{R}}^+ / \{0\}$.

In accordance with Appendix \ref{SAE-no-acotados}, we must now
determine the subspaces ${\cal K}_\pm$ by looking for linearly
independent eigenfunctions of the operator $H^{\dagger}_l$
corresponding to the eigenvalues $\pm i$, $\psi^{\pm}_{(x)}$.
Taking into account eq.\ (\ref{Hldagger}), it is easily seen from
\begin{equation}\label{Kpm}
H_l^\dagger \psi^{\pm}_{(x)}=\pm i \psi^{\pm}_{(x)}
\end{equation}
that the first derivative of $\psi^{\pm}_{(x)}$ is absolutely
continuous, as well as its derivatives of all order. Thus,
$\psi^{\pm}_{(x)} \in {\cal C}^\infty \bigcap L_2(\textbf{{R}}^+,2\pi
x\, dx)$, and the eigenvalue problem eq.\ (\ref{Kpm}) reduces to
a classical ordinary differential equation.

\bigskip

Then, eq.\ (\ref{Kpm}) leads to the following system of coupled
differential equations for the components, $\phi_{\pm}$ and
$\chi_{\pm}$, of the eigenfunctions $\psi^{\pm}$:
\begin{eqnarray}
    i\frac{d\chi_{\pm}}{dx}-i\left(\frac{\alpha-1}{x}+x\right)\chi_{\pm}
    =(\pm i-m)\phi_{\pm},\label{sistema-1}\\
    i\frac{d\phi_{\pm}}{dx}+i\left(\frac{\alpha}{x}+x\right)\phi_{\pm}
    =(\pm i+m)\chi_{\pm}.\label{sistema-2}
\end{eqnarray}
Replacing $\chi_{\pm}$ from eq.\ (\ref{sistema-2}) in eq.\
(\ref{sistema-1}), we get for the other component
\begin{equation}\label{phi}
  \phi_{\pm}''+\frac{1}{x}\phi_{\pm}'-\left(\left(\frac{\alpha}{x}\right)^{2}+x^{2}-
  2(1-\alpha)+m^2+1\right)\phi_{\pm}=0.
\end{equation}
Making the substitution
\begin{equation}\label{phi2}
  \phi_{\pm}=e^{-\frac{x^{2}}{2}}x^{-\alpha}F(x^{2}),
    \end{equation}
we obtain Kummer's equation \cite{A-S}
\begin{equation}\label{kum2}
x^2\frac{d^2F}{d(x^2)^2}(x^2)+\left[b-x^2\right]\frac{dF}{d(x^2)}(x^2)-
a\,F(x^2)=0,
\end{equation}
with $a=\frac{m^2+1}{4}>0$ and $b=1-\alpha = 1 - \kappa +l$.

This equation has two linearly independent solutions \cite{A-S},
$M(a,b,x^2)$ and $U(a,b,x^2)$, only the later of which leads to
$\phi_{\pm}\in L_2((\delta,\infty),2\pi x\, dx)$, with $\delta>0$.
On the other hand, the condition $\phi_{\pm}\in
L_2((0,\delta),2\pi x\, dx)$ requires $0<b<2$ (see \cite{A-S},
pag.\ 508).

Moreover, the condition that the second component (determined by
eq.\ (\ref{sistema-2})) satisfies $\chi_{\pm}\in
L_2(\textbf{R}^+,2\pi x\, dx)$, imposes $0<b<1$. This
requires\footnote{\label{no-entero} Notice that if $\kappa \in
\textbf{Z}$, the presence of the singular flux through the origin
amounts to a shift in the value of the orbital angular momentum
(as can be seen from eq.\ (\ref{ham})), without any further
consequence, $H_l$ being essentially self-adjoint. For brevity,
we will not further consider this case in what follows.} that
$\kappa \notin \textbf{Z}$ , and selects the subspace for which
$l$ is the integer part of $\kappa$, $\kappa - 1< l <\kappa$, as
the only one where nontrivial self-adjoint extensions exist.

Thus, for $l\neq [\kappa]$, $H_l$ is essentially self-adjoint,
admitting a unique self-adjoint extension given by the closure of
its graph (see Appendix \ref{clausura}).

On the other hand, for $l = [\kappa]$ we have found
one-dimensional subspaces ${\cal K}_\pm$, generated by the
solutions of eq.\ (\ref{Kpm}), $\psi^\pm$, given in components by
\begin{eqnarray}\label{psi-mas-menos}
  \phi_{\pm}=e^{-\frac{x^{2}}{2}}x^{-\alpha}U(a=\frac{m^2+1}{4}; b=1-\alpha;
 x^{2}), \\
 \chi_{\pm}=
 \left[\frac{-im\mp 1}{2}\right] \times  \nonumber \\
e^{-\frac{x^{2}}{2}}x^{-\alpha+1}U(a=\frac{m^2+5}{4};
 b=2-\alpha; x^2).
\end{eqnarray}
Therefore, $n_+ = 1 = n_-$, and $H_{[\kappa]}$ admits a
one-parameter ($\gamma$) family of (essentially) self-adjoint
extensions \cite{R-S}, $H_{[\kappa]}^\gamma$, which, as explained
in the Appendix \ref{SAE-no-acotados}, are in a one-to-one
correspondence with the isometries ${\cal U}_\gamma$ from ${\cal
K}_+$ onto ${\cal K}_-$:
\begin{equation}\label{isometria}
    {\cal U}_\gamma \psi^+ = e^{i \gamma} \psi^-,
\end{equation}
with $-\pi < \gamma \leq \pi$.

The functions $\psi_{(x)}$ in the domain of $H_{[\kappa]}^\gamma$
are of the form
\begin{equation}\label{psi-dom}
  \psi=\psi_0+c(\psi^+ + e^{i\gamma}\psi^-),
\end{equation}
where $\psi_0 \in C^{\infty}_{0}(\textbf{R}^+)$ and $c \in
\textbf{C}$, the action of $H_{[\kappa]}^\gamma$ being defined by
\begin{equation}\label{H-psi-dom}
  H_{[\kappa]}^\gamma \psi \equiv H_{[\kappa]} \psi_0 + c\,i(\psi^+ -
  e^{i\gamma}\psi^-).
\end{equation}

In Appendix \ref{clausura}, it is shown that the functions in the
closure of the graph of $H_{[\kappa]}$ are continuous and
vanishing for $x\rightarrow 0^+$. Therefore, the behavior at the
origin of the functions in the domain of the closure of
$H_{[\kappa]}^\gamma$, ${\cal
D}(\overline{H_{[\kappa]}^\gamma})$, is determined by the
behavior of $\psi^{(\gamma)} \equiv \psi^+ +  e^{i\gamma}\psi^-$,
whose components satisfy
\begin{eqnarray}\label{cc}
    \phi^{(\gamma)}=
    (1+e^{i\gamma})
    \frac{\Gamma (\alpha)}{\Gamma (\alpha+\frac{m^2+1}{4})}\ x^{-\alpha} +
    O(x^{\alpha}) \\ \label{cc1}
    \chi^{(\gamma)}=
    \frac{-i}{2}\left[m(1+e^{i\gamma})-i(1-e^{i\gamma})\right]
    \frac{\Gamma (1-\alpha)}{\Gamma
    (\frac{m^2+5}{4})}\times \\
    x^{-1+\alpha} +O(x^{1-\alpha}). \nonumber
\end{eqnarray}

This allows for the following characterization of the {\it
boundary conditions} the functions $\psi=\left( \begin{array}{c}
  \phi   \\
  \chi
\end{array}\right) \in {\cal D}(\overline{H_{[\kappa]}^\gamma})$ satisfy:
\begin{equation}\label{BBCC}
 \lim_{x\rightarrow
  0^+}\left\{x\left[ {\phi}\,{\chi^{(\gamma)}}-{\chi}\,
{\phi^{(\gamma)}}\right]\right\}= 0.
\end{equation}

We will use this condition in the next section to determine the
spectrum of $\overline{H_{[\kappa]}^\gamma}$.

\section{Spectrum of $\overline{H_{[\kappa]}^\gamma}$}\label{espectro-k}

In this section, making use of the {\it boundary condition}
deduced in eq.\ (\ref{BBCC}), we will determine the
eigenfunctions and eigenvalues of
$\overline{H_{[\kappa]}^\gamma}$. So we must solve the eigenvalue
problem
\begin{equation}\label{autov-ham}
    \overline{H_{[\kappa]}^\gamma} \psi = \lambda \psi.
\end{equation}

Notice that, since $\overline{H_{[\kappa]}^\gamma}$ is the
restriction of $H_{[\kappa]}^{\dagger}$ to
$\cal{D}(\overline{H_{[\kappa]}^\gamma)}\subset
\cal{D}(\overline{H_{[\kappa]}^\dagger)}$, both operators are
realized by the same differential operator (given in eq.\
(\ref{ham}), with $l$ replaced by $[\kappa]$). On the basis of an
argument similar to the one following eq.\ (\ref{Kpm}), we
conclude that we are looking for ${\cal C}^{\infty}$ solutions of
an ordinary differential equation. In terms of the components
$\phi$ and $\chi$, we get the pair of coupled differential
equations
\begin{eqnarray}\label{sistema}
    i\chi'-i\left(\frac{\alpha-1}{x}+x\right)\chi=(\lambda-m)\phi,\\
    i\phi'+i\left(\frac{\alpha}{x}+x\right)\phi=(\lambda+m)\chi.
    \label{sistema-chi}
\end{eqnarray}

Once again, the substitution given in eq.\ (\ref{phi2}) (now with
$0<\alpha<1$), leads to Kummer's equation for $F(x^2)$, eq.\
(\ref{kum2}), with $a=\frac{m^2-\lambda^2}{4}$ and $b=1-\alpha$.
The requirement that $\phi$ and $\chi$ belong to
$L_2(\textbf{R}^+,2\pi x\, dx)$ selects as the unique solution:
\begin{eqnarray}
    \phi_{\lambda}=
    e^{-x^2/2}x^{-\alpha}U(a=(m^2-\lambda^2)/4;b=1-\alpha;x^2)\\
    \chi_{\lambda}=
    \frac{i}{2}\left(\lambda-m\right)
    e^{-x^2/2}x^{1-\alpha}\times \\
    U(a=1+(m^2-\lambda^2)/4;b=2-\alpha;x^2),\nonumber
\end{eqnarray}
behaving, for $x\rightarrow 0^+$, as
\begin{eqnarray}\label{cc2}
    \phi_{\lambda}=
    x^{-\alpha} \frac{\Gamma (\alpha)}
    {\Gamma (\alpha+\frac{m^2-\lambda^2}{4})}+O(x^{\alpha}),\\
    \chi_{\lambda}=
    \frac{i(\lambda-m)}{2}
    x^{-1+\alpha}\frac{\Gamma (1-\alpha)}
    {\Gamma (1+\frac{m^2-\lambda^2}{4})}+O(x^{1-\alpha}).
\end{eqnarray}

So, the condition expressed in eq.\ (\ref{BBCC}) implies
\begin{eqnarray}\label{CCBB-autovector}
    \frac{\Gamma (\alpha)}
    {\Gamma (\alpha+\frac{m^2-\lambda^2}{4})}
    \left[m(1+e^{i\gamma})-i(1-e^{i\gamma})\right]
    \frac{\Gamma (1-\alpha)}{\Gamma
    (\frac{m^2+5}{4})}= \nonumber \\
   -(\lambda-m)
    \frac{\Gamma (1-\alpha)}
    {\Gamma (1+\frac{m^2-\lambda^2}{4})}
    (1+e^{i\gamma})
    \frac{\Gamma (\alpha)}{\Gamma
    (\alpha+\frac{m^2+1}{4})},
\end{eqnarray}
which can also be written as
\begin{eqnarray}\label{energ}
    G(\lambda)\equiv(\lambda-m)\left(\frac
    {\Gamma(\alpha+m^2/4-\lambda^2/4)}{\Gamma(1+m^2/4-\lambda^2/4)}\right)
    = \nonumber \\
    (\tan{(\gamma/2)}-m)\left(\frac
    {\Gamma(\alpha+m^2/4+1/4)}{\Gamma(m^2/4+5/4)}\right)\equiv \beta(\gamma).
\end{eqnarray}
This is a transcendental equation determining the eigenvalues of
$\overline{H_{[\kappa]}^\gamma}$. The whole dependence on
$\lambda$ is contained in $G(\lambda)$, on the l.h.s. This
function has simple zeros at $\lambda = m$ and
$\lambda=\pm\sqrt{4+m^2+4n}$, and simple poles at
$\lambda=\pm\sqrt{4\alpha+m^2+4n}$, for $n=0,1,2,...$ (see Figure
1).

On the r.h.s.\ of eq.\ (\ref{energ}), $\beta(\gamma)$ is a
constant depending only on $m$, $\alpha$ and the parameter
$\gamma$ characterizing the self-adjoint extension of
$H_{[\kappa]}$. It can take all real values with $\gamma$ ranging
from $-\pi$ to $\pi$, being $\beta(\gamma)>0$ for
$\gamma_0<\gamma < \pi$, and $\beta(\gamma)<0$ for
$-\pi<\gamma<\gamma_0$, where $\gamma_0=2\arctan(m)$.
\begin{figure}\label{F1}
    \epsffile{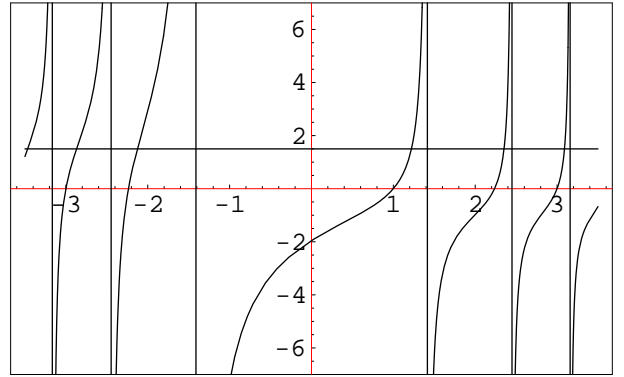}
    \caption{Graphics of $G(\lambda)$ for $m=1$ and $\alpha=1/4$.
The horizontal line corresponds to a positive value of $\beta(\gamma)$.}
   \end{figure}

It is evident from Figure 1.\ that the spectrum of
$\overline{H_{[\kappa]}^\gamma}$ does depend on $\gamma$. If
$\gamma_0<\gamma<\pi$, the eigenvalues lie between a zero of
$G(\lambda)$ and the nearest pole on its right: For $\lambda>m$
\begin{equation}\label{la-pos}
   \sqrt{m^2+4(N+1)}<\lambda_{N}<\sqrt{m^2+4(\alpha+N+1)},
\end{equation}
with $N=0,1,2,\ldots$ and, for $\lambda<m$,
\begin{equation}\label{la-neg}
  -\sqrt{m^2-4N}<\lambda_{N}<-\sqrt{m^2+4(\alpha-N-1)},
\end{equation}
with $N=-1,-2,-3,\ldots$

For $-\pi<\gamma<\gamma_0$ the eigenvalues are bounded on the left
by a pole and on the right by the nearest zero of $G(\lambda)$:
For $\lambda>m$
\begin{equation}\label{la-pos-neg}
    \sqrt{m^2+4(\alpha+N-1)}<\lambda_{N}<\sqrt{m^2+4 N},
\end{equation}
with $N=1,2,3,\ldots$,
\begin{equation}\label{la-0}
    -\sqrt{m^2+4\alpha}<\lambda_{0}<m,
\end{equation}
 and, for $\lambda<0$,
\begin{equation}\label{la-neg-neg}
   -\sqrt{m^2+4(\alpha-N)}<\lambda_{N}<-\sqrt{m^2-4N}
\end{equation}
with $N=-1,-2,-3,\ldots$

Notice that there is only one level with $|\lambda|<
\sqrt{m^2+4\alpha}$. Moreover, the spectrum of
$\overline{H_{[\kappa]}^\gamma}$ is symmetric with respect to the
origin only for $\gamma=\pi$ and (except for the eigenvalue
$\lambda_0=m$) for $\gamma=\gamma_0$.

\section{Spectrum of $\overline{H_{l}}$ for $l\neq [\kappa]$}

In this section we complete the description of the Hamiltonian
spectrum by computing the eigenfunctions and eigenvalues of
$\overline{H_{l}}$ for $l\neq [\kappa]$.

As we saw in Section \ref{Section-SAE}, in the present case $H_l$
is essentially self-adjoint, admitting a unique self-adjoint
extension given by the closure of its graph. According to
Appendix \ref{clausura}, the vectors in ${\cal
D}(\overline{H_l})$ are absolutely continuous functions vanishing
at the origin.

We are looking for solutions of the system given by eqs.\
(\ref{sistema}-\ref{sistema-chi}) in this domain. Once again, by
an argument similar to the one employed in Section
\ref{Section-SAE}, one can see that the eigenvectors belong to
${\cal C}^\infty \bigcap L_2(\textbf{R}^+,2\pi x\, dx)$.

Following the same steps as in Section \ref{espectro-k}, one
obtains the solutions in terms of Kummer's functions. It is
convenient to write them in terms of the following pair of
linearly independent solutions of eq.\ (\ref{kum2}):
\begin{eqnarray}\label{autof-no-k}
 F_1(x^2)=M(a;b;x^2),\\
 F_2(x^2)=x^{2\alpha}M(1+a-b;2-b;x^2),
\end{eqnarray}
where $a=\frac{m^2-\lambda^2}{4}$ and $b=1-\alpha$, with
$\alpha=\kappa - l$ ($\notin \textbf{Z}$ - see footnote
\ref{no-entero} -). We will consider the cases $l<[\kappa]$ and
$l>[\kappa]$ separately.


\subsection*{\underline{$l<[\kappa]$}}

 For $l<[\kappa]$ ($\alpha=\kappa-l>1$), only $F_2(x^2)$ leads to
 functions
\begin{eqnarray}\label{phi-no-k-men}
  \phi_{\lambda}=
    e^{-x^2/2}x^{\alpha}M(\frac{m^2-\lambda^2}{4}+\alpha;1+\alpha;x^2), \\
    \chi_{\lambda}=
    \left[\frac{2i}{(m + \lambda)}\right]
    e^{-x^2/2}x^{-1+\alpha}\times \nonumber\\
    \left[\alpha
    M(\frac{m^2-\lambda^2}{4}+\alpha;1+\alpha;x^2)+\right. \nonumber\\
    \left.\frac{m^2-\lambda^2+4\alpha}{4(1+\alpha)}x^2
    M(\frac{m^2-\lambda^2}{4}+\alpha+1;2+\alpha;x^2)\right],
\end{eqnarray}
which are in $L_2((0,\delta>0),2\pi x\,dx)$. Moreover, the
condition $\phi_{\lambda},\chi_{\lambda}\in
L_2((\delta,\infty),2\pi x\,dx)$ requires that $M(1+a-b;2-b;x^2)$
reduces to a polynomial, which occurs only when
\begin{equation}\label{cond-autovalor-caso1}
  1+a-b=\kappa - l + \frac{m^2-\lambda^2}{4} = -n,
\end{equation}
with $n=0,1,2,\ldots$ \ So, the eigenvalues are given by
\begin{equation}\label{autovalor-caso1}
  \lambda=\pm \, 2\sqrt{m^2/4+\kappa +N}, \ N=-l,-l+1,-l+2,\ldots
\end{equation}
Notice that both, the eigenfunctions and eigenvalues depend on
the singular flux $\kappa$.


\subsection*{\underline{$l>[\kappa]$}}

For $l>[\kappa]$ ($\alpha=\kappa-l<0$), only $F_1(x^2)$ leads to
functions
\begin{eqnarray}
    \phi_{\lambda}=
    e^{-x^2/2}x^{-\alpha}M(\frac{m^2-\lambda^2}{4};1-\alpha;x^2),\\
    \chi_{\lambda}=
    \left[\frac{i(m-\lambda)}{2(1-\alpha)}\right]
    e^{-x^2/2}x^{1-\alpha}\times \nonumber\\
    M(\frac{m^2-\lambda^2}{4}+1;2-\alpha;x^2),
\end{eqnarray}
which are in $L_2((0,\delta>0),2\pi x\,dx)$. Once again, the
condition $\phi_{\lambda},\chi_{\lambda}\in
L_2((\delta,\infty),2\pi x\,dx)$ requires that $M(a;b;x^2)$
reduces to a polynomial, which now occurs when
\begin{equation}\label{cond-autovalor-caso2}
  a=\frac{m^2-\lambda^2}{4} = -n,
\end{equation}
with $n=0,1,2,\ldots$ \ This time, the eigenvalues are given by
\begin{equation}\label{autovalor-caso2}
  \lambda=\pm \, 2\sqrt{m^2/4+N}, \ N=0,1,2,\ldots
\end{equation}
In the present case the eigenfunctions do depend on the singular
flux, but the eigenvalues are independent of $\kappa$.

\bigskip

Finally, notice that in both cases ($l<[\kappa]$ and $l>[\kappa]$)
the eigenfunctions obtained vanish at the origin, thus belonging
to the domains ${\cal D}(\overline{H_l})$ of the corresponding
operator.

\appendix

\section{Self-adjoint extensions of unbounded
operators}\label{SAE-no-acotados}

In this Appendix we briefly review the theory of deficiency
indices of von Neumann (for an extended presentation of the
subject, see Ref.\ \cite{R-S}). We first recall the definition of
the adjoint of a given linear operator.

Let $A$ be a linear operator defined on a dense subspace ${\cal
D}(A)$ of a Hilbert space $H$. The domain of definition of the
adjoint operator $A^{\dagger}$, ${\cal D}(A^{\dagger})$, is the
set of vectors $\psi \in H$ making the inner product
$(\psi,A\phi)$ continuous in $\phi \in {\cal D}(A)$. In virtue of
the Riesz - Fischer theorem, for any such $\psi$ there exists a
unique vector $\chi\in H$ satisfying $(\psi,A\phi)=(\chi,\phi), \
\forall \phi \in {\cal D}(A)$. One defines $A^{\dagger}\psi
\equiv \chi$.

A linear operator $A$ is symmetric if
\begin{equation}\label{symmetric}
  \left( \phi_1,A\, \phi_2\right)=\left( A\,\phi_1, \phi_2\right),
  \forall \phi_1,\phi_2 \in {\cal D}(A).
\end{equation}
A linear operator $A$ is self-adjoint if it coincides with its
adjoint $A^{\dagger}$, i.e. if ${\cal D}(A^{\dagger})={\cal
D}(A)$ and
\begin{equation}\label{autoadjunto}
A^{\dagger}\phi = A \phi, \forall \phi \in {\cal D}(A).
\end{equation}

To establish the conditions a closed\footnote{Recall that an
operator is closed if its graph is a closed subset of $H\times
H$. Every symmetric operator defined on a dense set is closable,
i.e., has a closed symmetric extension.} symmetric operator must
satisfy to be self-adjoint, a few definitions are in order. Let
${\cal K}_{\pm}= Ker(A^{\dagger}\mp i)$ be the characteristic
subspaces of $A^{\dagger}$ corresponding to the $\pm i$
eigenvalues respectively. The {\it deficiency indices} of the
operator $A$, $n_{\pm}$, are defined as the dimensions of the
subspaces ${\cal K}_{\pm}$.

It is worth recalling that a closed symmetric operator is
self-adjoint if and only if its deficiency indices are zero
\cite{R-S}. However, if the deficiency indices are not zero but
equal the operator admits a family of self-adjoint extensions
whose construction can be carried out by means of the following
theorem \cite{R-S}: {\it  Let $A$ be a closed symmetric operator
whose deficiency indices $n_{\pm}$ are equal; then it admits a family
of self-adjoint extensions which are in a one-to-one
correspondence with the unitary maps from ${\cal K}_+$ onto
${\cal K}_-$.}

In fact, let ${\cal U}$ be such an isometry, then the
corresponding self-adjoint extension $A_{\cal U}$ has domain
${\cal D}(A_{\cal U})=\{\psi: \psi = \phi + \phi_+ + {\cal
U}(\phi_+)\}$, where  $\phi \in {\cal D}(A)$, and $\phi_+ \in
{\cal K}_+$. The action of the extension $A_{\cal U}$ is given by
\begin{equation}\label{accion-de-extension}
  A_{\cal
U}(\phi + \phi_+ + {\cal U}(\phi_+)) = A(\phi) + i \phi_+ - i
{\cal U}(\phi_+).
\end{equation}

This provides a method for constructing the self-adjoint
extensions of closed symmetric operators with equal deficiency
indices by identifying each possible unitary map from ${\cal K}_+$
onto ${\cal K}_-$.

\section{Closure of ${H_{l}}$}\label{clausura}

In this appendix we will study the  closure $\overline{H_{l}}$ of
the operator in eq.\ (\ref {ham}),
\begin{equation}\label{app-ham}
  H_{l}=
\left(  \begin{array}{cc}
    m & i\left(\frac{d}{dx}+\frac{1-\alpha}{x}-x\right) \\
    i\left(\frac{d}{dx}+\frac{\alpha}{x}+x\right) & -m \
  \end{array}\right),
\end{equation}
defined on ${\cal D}(H_{l})=C^{\infty}_{0}(\textbf{R}^+)$, a
dense subspace of $L_2(\textbf{{R}}^+,2\pi x\, dx)$. It will be
shown that the functions in the domain of definition of
$\overline{H_{l}}$ are continuous near the origin, and vanishing
for $x \rightarrow 0^+$.

In order to obtain ${\cal D}(\overline{H_{l}})$ we must add to
the domain of $H_l$ the limit points of the Cauchy sequences in
${\cal D}(H_l)$ whose images by $H_l$ are also Cauchy sequences.

So, let us consider a Cauchy sequence $\{\psi_n\}_{n\in
\mathbf{N}}$ with $\psi_n \in {\cal D}(H_l),\forall n \in
\mathbf{N}$, and such that $\{H_l \psi_n\}_{n\in \mathbf{N}}$ is
also a Cauchy sequence. Therefore, given $\varepsilon>0$,
\begin{eqnarray}\label{sequence1}
\| \psi_n - \psi_m \|^2 < \varepsilon \\ \label{sequence2}
 \|H_l(\psi_n - \psi_m)\|^2 < \varepsilon
\end{eqnarray}
for $n,m$ sufficiently large. Making use of eq.\ (\ref{app-ham}),
it is easily seen that
\begin{eqnarray}\label{norma-Hn}
\|H_l(\psi_n - \psi_m) \|^2= \nonumber \\
    \int_{0}^{\infty}\left(|\phi' |^2 + p_{(x)}|\phi|^2+
    |\chi' |^2 + q_{(x)}|\chi|^2\right)2\pi xdx,
\end{eqnarray}
where we have denoted by $\phi$ and $\chi$ respectively the upper
and lower component of $(\psi_n - \psi_m)$, while the functions
$p(x),q(x)$ are given by
\begin{eqnarray}
    p{(x)}=\left[\left(\frac{\alpha}{x}+x\right)^2+m^2-2\right],\\
    q{(x)}=\left[\left(\frac{1-\alpha}{x}+x\right)^2+m^2+2\right],
\end{eqnarray}
and are $O(x^{-2})$ for $x\rightarrow 0^+$ (since we are taking
$\alpha \notin \textbf{Z}$ - see footnote \ref{no-entero} -). It
is not hard to see that both $p(x)$ and $q(x)$ are positive in the
interval $[0,\delta]$ for some positive $\delta$. Only $p(x)$ can
change its sign in an interval $(x_1,x_2)$ (depending on $\alpha$
and $m$), with $0<\delta<x_1<x_2<\infty$. Notice that the
integrand of eq.\ (\ref{norma-Hn}) (obtained through an
integration by parts) could take negative values only in
$(x_1,x_2)$, as a consequence of the term $p(x)|\phi(x)|^2$.

Moreover, for $\delta$ small enough, we can choose $K>0$ such that
$p(x)>{K}/{x^2}$.  Taking into account eqs.\ (\ref{sequence1})
and (\ref{norma-Hn}), for a given $\varepsilon>0$, we can write
\begin{eqnarray}\label{cota-phi-phi'}
    \int_{0}^{\delta}|\phi' |^2 x\,dx <\varepsilon, \
    \int_{0}^{\delta}\frac{|\phi|^2}{x}\, dx < \varepsilon,
\end{eqnarray}
and
\begin{eqnarray}\label{cota-chi-chi'}
    \int_{0}^{\delta}|\chi' |^2 x\,dx <\varepsilon, \
    \int_{0}^{\delta}\frac{|\chi|^2}{x}\, dx < \varepsilon,
\end{eqnarray}
if $n,m$ are large enough. Therefore,
\begin{equation}\label{sec-fund}
  \{\sqrt{x}\, \psi'_n(x) \} \ {\rm and} \ \{ \psi_n(x)/ \sqrt{x}\}
\end{equation}
are Cauchy sequences in $L_2(0,\delta)$ (with respect to the
usual Lebesgue measure), as well as the sum
\begin{equation}\label{sec-fund-der}
  \{\sqrt{x}\, \psi'_n(x) + \psi_n(x)/ (2\sqrt{x})\} =\{ \left[\sqrt{x}\,
  \psi_n(x)\right]'\}.
\end{equation}

Let us call $\Phi(x)=\lim_{n \rightarrow \infty}\left[\sqrt{x}\,
\psi_n(x)\right]' \in L_2(0,\delta)$, and denote its primitive by
\begin{equation}\label{primitive}
  \sqrt{x}\, \Psi(x)\equiv \int_0^x \Phi(y)\, dy,
\end{equation}
which is an absolutely continuous function \cite{R-S} in
$(0,\delta)$. In particular, $\Psi(x)$ is continuous in
$(0,\delta)$.

On the basis of
\begin{eqnarray}\label{conv-unif}
    |\sqrt{x}(\Psi(x)-\psi_n(x))|=\nonumber\\
    \left|\int_0^x\left[\Phi(y)-(\sqrt{y}\,\psi_n(y))'\right]dy\right|\leq
    \nonumber \\
    \sqrt{\int_{0}^{\delta}\left|\Phi(y)-(\sqrt{y} \psi_n(y))'\right|^2 dy}
    \sqrt{\int_0^\delta 1\,dy} \nonumber \\
   \longrightarrow 0, \  {\rm for} \ n \rightarrow \infty,
\end{eqnarray}
we conclude that the sequence $\{\sqrt{x}\,\psi_n(x)\}$ converges
uniformly to $\sqrt{x}\,\Psi(x)$ in $(0,\delta)$, and
consequently also in the metric of $L_2(0,\delta)$,
\begin{equation}\label{lim2}
    \lim_{n\rightarrow
    \infty}\left\{\sqrt{x}\,\psi_n(x)\right\}=\sqrt{x}\,\Psi(x).
\end{equation}
(Notice that $\Psi(x)$ is the limit of $\{\psi_n(x)\}$ in
$L_2[(0,\delta),x\,dx]$.)

In addition, it is straightforward to show that
\begin{eqnarray}
    \frac{\Psi(x)}{\sqrt{x}}=\lim_{n\rightarrow
    \infty}\left\{\frac{\psi_n(x)}{\sqrt{x}}\right\}.\label{lim22}
\end{eqnarray}
Then, we conclude from eqs.\ (\ref{lim22}) and (\ref{primitive})
that
\begin{eqnarray}\label{lim3}
  \lim_{n\rightarrow\infty}\left\{\sqrt{x}\psi'_n(x)\right\}=\sqrt{x}\Psi'(x),
\end{eqnarray}
in the metric of $L_2(0,\delta)$.

Therefore, we can write
\begin{equation}\label{bounds}
    \int_{0}^{\delta}|\Psi' |^2 x\,dx <\infty, \
    \int_{0}^{\delta}\frac{|\Psi|^2}{x}\, dx < \infty.
\end{equation}
This implies that $\Psi'(x)\cdot\Psi(x)=1/2\left(\Psi(x)\cdot
\Psi(x) \right)' \in L_1(0,\delta)$.

On the other hand, the components of $\Psi(x)$, $\Psi_\alpha(x)$
with $\alpha = 1,2$, are absolutely continuous functions in
$(\epsilon,\delta)$, for $\epsilon<\delta$, by virtue of eq.\
(\ref{primitive}). In consequence
\begin{equation}\label{barrow}
  \int_{\epsilon}^{\delta}\left(\Psi_\alpha^2(x)\right)'\,dx=
  \Psi_\alpha^2(\delta)-\Psi_\alpha^2(\epsilon).
\end{equation}
In this expression we can take the $\lim_{\epsilon\rightarrow
0^+}$, proving that the continuous function $\Psi_\alpha^2(x)$
has a well defined limit for $x\rightarrow 0^+$. Moreover, on
account of eq.\ ({\ref{bounds}}), this limit must be zero.

\bigskip

As a consequence of the previous results, we conclude that the
behavior near the origin of the functions in ${\cal
D}(\overline{H_{[\kappa]}^\gamma})$ is dominated by the functions
in ${\cal K }_\pm$ (see eqs.\ (\ref{cc}-\ref{cc1})).

On the other hand, since the restriction of the Hamiltonian to the
subspaces with $l\neq [\kappa]$ is, as already mentioned,
essentially self-adjoint, the behavior of the functions at the
origin is dictated by its closure, therefore being continuous and
satisfying the boundary condition
\begin{equation}\label{limit}
  \lim_{x\rightarrow 0^+}\psi(x)=0.
\end{equation}

\bigskip

{\bf Acknowledgements:} The authors thank M.A. Muschietti and
E.M.\ Santangelo for useful discussions and comments. H.\ F.\
thanks P.\ Giacconi and R.\ Soldati for calling his attention to
this problem. This work was partially supported by ANPCyT (PICT'97
Nr.\ 00039), and CONICET (PIP Nr.\ 0459) of Argentina.

\end{document}